**Ponderomotive generation and detection of attosecond free-electron pulse trains**

M. Kozák[1,2,*], N. Schönenberger[1], and P. Hommelhoff[1]

[1] *Department of Physics, Friedrich-Alexander-Universität Erlangen-Nürnberg (FAU), Staudtstrasse 1, 91058 Erlangen, Germany, EU*

[2] *Faculty of Mathematics and Physics, Charles University, Ke Karlovu 3, 12116 Prague 2, Czech Republic, EU*

[*]Corresponding author: martin.kozak@fau.de

**Abstract**

Atomic motion dynamics during structural changes or chemical reactions have been visualized by picosecond and femtosecond pulsed electron beams via ultrafast electron diffraction and microscopy. Imaging the even faster dynamics of electrons in atoms, molecules and solids requires electron pulses with sub-femtosecond durations. We demonstrate here the all-optical generation of trains of attosecond free-electron pulses. The concept is based on the periodic energy modulation of a pulsed electron beam via an inelastic interaction with the ponderomotive potential of an optical travelling wave generated by two femtosecond laser pulses at different frequencies in vacuum. The subsequent dispersive propagation leads to a compression of the electrons and the formation of ultrashort pulses. The longitudinal phase space evolution of the electrons after compression is mapped by a second phase-locked interaction. The comparison of measured and calculated spectrograms reveals the attosecond temporal structure of the compressed electron pulse trains with individual pulse durations of less than 300 as. This technique can be utilized for tailoring and initial characterization of sub-optical cycle free-electron pulses at high repetition rates for stroboscopic time-resolved experiments with sub-femtosecond time resolution.

**Main text**

The natural time scales of electron wavepacket dynamics in atoms, molecules and solids lie in the attosecond time domain (1 as=$10^{-18}$ s). The observation of such ultrafast phenomena in various systems has recently been enabled due to the rapid development of attosecond science and metrology [1-3]. The extreme ultraviolet (XUV) attosecond pulses serve for, e.g., the direct characterization of the temporal evolution of the electric field of optical pulses [4,5], for studying the ultrafast response of solid-state materials [6,7] or for time-resolved photoemission studies [8]. However, only a collective response of many atoms and molecules has been examined so far due to the lack of the high spatial resolution required to address individual quantum systems with a typical size of less than 1 nm. The experimental spatial resolution is limited by diffraction (even XUV attosecond pulses with a photon energy of 100 eV have a wavelength of 10 nm), by the transverse coherence of the XUV beams and the quality of XUV optics. On the other hand, atomic spatial resolution is commonly provided by electron beams with keV energies (typically 1-300 keV) with de Broglie wavelengths of $\lambda_{dB}$=1-40 pm employed in electron microscopy and diffraction.

A combination of atomic spatial and femtosecond temporal resolutions has been reached in ultrafast electron diffraction (UED) and microscopy (UEM) experiments, which have already provided insights to dynamics of chemical reactions [9-11], phase transitions [12,13] or ultrafast switching of magnetic domains [14,15]. Unlike optical or XUV pulses, the propagation of electron pulses in vacuum is dispersive, preventing attosecond pulse durations to be directly reached by controlling the electron emission time on sub-optical cycle time scales [16]. The achievable temporal resolution of UED and UEM in a single-electron per pulse regime is so far limited to ~30 femtoseconds by dispersive broadening of electron pulses and/or by the timing jitter of the electron pulse compression and gating schemes [17-20].

Attosecond timing control can be transferred from the optical domain to electrons via a coherent interaction between electron wavepackets accelerated to keV-MeV energies and light. Two physical principles are feasible for this application, both based on introducing a time-dependent energy modulation to the propagating electrons and subsequent ballistic compression due to dispersion of the electrons in vacuum. The first technique rests on inelastic electron scattering at optical near-fields generated on the surface of various nanostructures [21-23]. Coherent, sub-optical-cycle control of electrons has been demonstrated via this technique [23,24], and the generation of attosecond electron pulse trains is feasible [25,26]. The second method, proposed in [27-29] and demonstrated in this work, is based on the interaction of electrons with a ponderomotive potential of a co-propagating optical travelling wave generated by two optical fields at different frequencies in vacuum (without the presence of any structure close to the electron beam). In this scheme, the propagation velocity of the travelling wave is synchronized to the electron velocity, leading to the generation of an optical standing wave in the electrons´ rest frame. Due to the gradient of the optical intensity in the direction of the electron propagation, electrons are pushed out of the high intensity regions by the ponderomotive force. Consequently, their longitudinal momentum component is changed. This process was very recently experimentally demonstrated via the observation of a large broadening of post-interaction electron spectra [30].

Due to the absence of any structure in the laser focus, the strength and repetition rate of the ponderomotive compression is not limited by the optical damage threshold of solid state materials necessary for other optically-driven compression schemes [21-26]. Further, the ponderomotive technique can be advantageous at low electron energies, where the interaction with optical near-fields becomes inefficient and the transmission of the electrons through the compression element [25,26] limits the electron beam brightness. Last, the force acting on the electrons in the case of the interaction with optical near-fields has both a longitudinal and a transverse component that are phase shifted by π/2 and represent an obvious potential cause of spatial beam distortions. For the ponderomotive interaction, there is only the longitudinal force for on-axis electrons.

In this proof-of-concept paper we use the ponderomotive interaction with an optical travelling wave to both generate and detect the sub-cycle temporal structure in the longitudinal density of free electrons. In the slowly varying envelope approximation and with a small maximum relative energy spread of $\Delta E_{k,max}(t)/E_{k0}$ <5% leading to a negligible phase slippage of the electrons with respect to the travelling wave during the interaction, the impulse approximation can be applied to describe the final electron energy modulation. Using this simplified description, the electrons´ kinetic energy immediately after the interaction is modulated as $E_k = E_{k0} + \Delta E_{k,max}(t)\sin(2\pi t/T)$, where $E_{k0}$ is the initial electron energy, $\Delta E_{k,max}(t)$ is the temporal envelope of the energy modulation amplitude of the electrons and $T=2\pi/(\omega_1-\omega_2)=16.6$ fs is the time period of the travelling wave given by the difference frequency of the two pulsed laser beams (with angular frequencies $\omega_1$ and $\omega_2$) generating the optical travelling wave (Fig. 1(b)). During each period, a part of the electron pulse interacts with the time-dependent ponderomotive potential close to its minimum, where the potential can be approximated as parabolic in time [28]. As a consequence, these electrons obtain a linear energy chirp of $\kappa = dE_k/dt$. With the relative velocity change $\Delta v/v_0$<<1, the dispersion relation of electrons in vacuum $E_k(p)$ is close to linear, leading to a temporal compression of the linearly chirped part of the electron distribution during ballistic propagation and a formation of a series of attosecond electron pulses separated by $T$. To detect the generated attosecond time structure of the pulsed electron beam we monitor the evolution of the electron distribution in the longitudinal phase-space after the interaction. This is achieved using an interaction with a second spatio-temporally separated optical travelling wave that serves for time-to-energy mapping of the electron density with sub-cycle temporal resolution (Fig. 1). The ponderomotive potential of an individual travelling wave generated by two pulsed laser beams can be written as [28,30]:

$$U_p \cong 2\xi(t,t_0,z,z_0) \frac{e^2}{m_0(\omega_1+\omega_2)^2} \cos\left[(\omega_1-\omega_2)(t-t_0) - (\omega_1\cos\alpha - \omega_2\cos\beta)\frac{(z-z_0)}{c} + (\varphi_1-\varphi_2)\right]$$

. (1)

Here $\xi(t,t_0,z,z_0)$ is the intensity envelope function of the optical travelling wave temporally and spatially centered at $t_0$ and $z_0$, respectively, $e$ is the electron charge, $m_0$ the electron mass, $c$ the speed of light and $\varphi_1$ and $\varphi_2$ are the carrier-envelope phases of the two laser pulses. The electron beam propagates along the $z$-axis.

The two spatio-temporally separated optical travelling waves - the first for the attosecond electron pulse train generation and the second for its analysis - are generated using two independent Michelson interferometers (Fig. 1(a)). The relative phase difference of the two travelling waves $\Delta\varphi_{tw}$ is, according to Eq. (1), given by:

$$\Delta\varphi_{tw} = \Delta\varphi_2 - \Delta\varphi_1 = (\omega_1-\omega_2)(t_0^{(1)}-t_0^{(2)}) - (\omega_1\cos\alpha - \omega_2\cos\beta)\frac{(z_0^{(1)}-z_0^{(2)})}{c}, \quad (2)$$

where $\Delta\varphi_1$ and $\Delta\varphi_2$ are the relative phase differences introduced in the two Michelson interferometers and $t_0^{(1)}$, $t_0^{(2)}$, $z_0^{(1)}$ and $z_0^{(2)}$ are the temporal and spatial centers of the envelope functions $\xi^{(1)}(t,t_0^{(1)},z,z_0^{(1)})$ and $\xi^{(2)}(t,t_0^{(2)},z,z_0^{(2)})$ describing the intensity envelopes of the two separated travelling waves. We note that $\Delta\varphi_{tw}$ does not depend on the carrier-envelope phases $\varphi_1$ and $\varphi_2$ of the laser pulses.

The experiment is carried out using a femtosecond optical parametric amplifier producing the required pulses at frequencies $\omega_1=2\pi\nu_1=2\pi\times2.18\times10^{14}$ rad/s and $\omega_2=2\pi\times1.58\times10^{14}$ rad/s with FWHM pulse durations of $\tau_l\approx50$ fs. The angles of incidence of the two pairs of laser beams with respect to the pulsed electron beam at the initial energy of $E_{k0}=23.5$ keV are $\alpha=41°$ (beams $\omega_1^{(1)}$, $\omega_1^{(2)}$) and $\beta=107°$ ($\omega_2^{(1)}$, $\omega_2^{(2)}$). This choice of incidence angles leads to an optical travelling wave with the intensity fronts perpendicular to the propagation direction of the electrons. As a consequence, the only nonzero component of the ponderomotive force is in the longitudinal direction with respect to the electron beam propagation. The electron beam is focused to the transverse spot size (FWHM) of $w_e \approx 200$ nm, the initial electron pulse duration (FWHM) is $\tau_{e0}=460$ fs and pulse charge is $|q|<|e|$ to avoid pulse broadening by the space-charge effects. Because the transverse size of the electron beam is much smaller than the size of the laser foci, the energy modulation across each transverse plane of the electron beam is homogenous. The electron energy spectra after the interaction with both optical travelling waves are measured by a magnetic spectrometer and a microchannel plate detector (for details see [31]).

The measured electron energy spectra as a function of the relative phase $\Delta\varphi_{tw}$ (hereafter referred to as spectrograms) after the interaction with both optical travelling waves are plotted in Fig. 2(a) for different maximum values of the linear chirp $\kappa^{(1)}$ introduced by the first interaction. The spectrograms clearly reveal phase oscillations: The energy center of mass of the electron distribution (black lines in Fig. 2(a)) is modulated periodically with the relative phase between the two optical travelling waves. The experimental spectrograms are compared to numerical simulations (Fig. 2(b)) calculated using the measured values of the linear chirp introduced by the first interaction $\kappa^{(1)}$, the streaking power of the second interaction $\kappa^{(2)}$ (see Fig. S2 in [31]) and the drift distance of $d=41\pm2$ μm. Due to a mismatch between the initial electron pulse durations and the laser pulse durations $\tau_{e0}> \tau_l$, the amplitude of the electron energy modulation $\Delta E_{k,\max}(t) = \Delta E_{k,\max}(0)\xi(t)$ is given by the temporal envelope of the travelling wave $\xi(t)$. Consequently, the value of $\kappa$ as well as the drift distance $d_c = \left(\gamma^3 m_0 v_0^3\right)/\kappa$ between the first interaction and the temporal focus are not constant for electrons interacting with different periods of the travelling wave, leading to a complex temporal distribution of the electrons in the pulse train. Periodic attosecond pulse trains with constant duration can be generated by using longer laser pulses fulfilling the condition $\tau_l > \tau_{e0}$.

Because of the above mentioned experimental conditions, the temporal distributions of the electron density in the spatial center of the second interaction $z_0^{(2)}$ cannot be obtained directly from the measured spectrograms. Instead we use an indirect method, where the temporal evolutions of the electron density are calculated by integration of the electrons over the entire transverse electron beam profile from the distributions corresponding to the calculated spectrograms (results plotted in Fig. 2(b) and (c)). Optimal compression and creation of the attosecond electron pulse trains with the calculated duration of the shortest individual pulses of $\tau_e$=260 as (see Fig. 2(d)) is reached for $\kappa^{(1)}$=95 eV/fs. For the relatively broad window of $\kappa^{(1)}$=80-150 eV/fs, the center part of each pulse train (within the full-width at half-maximum of the temporal envelope of the laser pulse) consists of sub-femtosecond electron pulses. Results obtained with a longer drift distance of $d$=73±3 μm between the two interaction regions are plotted in Fig. S3 in [31].

The generated attosecond electron pulse trains contain both the bunched electrons and the electrons that interact with the first travelling wave outside of the time window of the parabolic potential needed for ballistic compression. These electrons are homogenously distributed in time at the site of the second interaction and show up as the background electron density between the pulses in the trains in Fig. 2(c). In the temporal focus ($\kappa^{(1)}$=95 eV/fs), the fraction of electrons present within a 1 fs window around the individual pulse maxima is approximately 30%. This number can be further improved by post-interaction spectral filtering of the electrons to 50%.

The state of the electrons in the longitudinal phase space after ballistic propagation is experimentally determined from the measured spectrograms using the difference between the phase of the energy minima and maxima $\Delta\varphi_{E\mathrm{max}\text{-}E\mathrm{min}}$ (dashed lines in Fig. 2(a) and (b)). In Fig.

3, the measured phase shift $\Delta\varphi_{E\text{max}-E\text{min}}$ is plotted as a function of $\kappa^{(1)}$ (points) in comparison with the simulation results (dashed line). Here the ideal compression corresponds to a phase shift of ∼3/4π. The slight difference between measurement and simulation data is due probably to a small deviation of the spatial and temporal shape of the pulsed laser beams from the Gaussian shape assumed in the simulations, or by a small imperfection of the spatio-temporal overlap of all four laser beams with respect to the electron pulses.

The ponderomotive interaction can also serve as a source of isolated attosecond electron pulses generated via a two-stage compression scheme (see Fig. S4 in [31]). The timing jitter of the first compression stage based on the interaction with RF or THz fields (∼10 fs for the RF [17] and ∼3 fs for THz [18]) can be compensated by the interaction with the phase-stable optical travelling wave allowing to reach sub-100 as temporal resolution in time-resolved experiments (see [31] for details).

The critical quantity for future applications of this scheme in time-resolved imaging with attosecond resolution is the brightness of the final compressed part of the electron beam. In our experiments, the average brightness is limited by the repetition rate $f_{\text{rep}}$=1 kHz of the laser system used. However, laser pulses with energies down to a few hundred nJ will be used in future, which will still provide an energy modulation of the order of 10 eV, sufficient for attosecond electron pulse train generation over a significantly longer drift distances of ∼1-5 mm. The repetition rate of the experiment thus will be increased to 100 kHz-10 MHz. The total current of the compressed part of the electron distribution (30% of the electrons) is then expected to reach the pA level.

Apart from applications in ultrafast electron diffraction or microscopy, the here demonstrated attosecond compression of electron pulses can by utilized in electron sources for a new generation of laser-driven particle accelerators [22] or as a tool for the manipulation of the longitudinal phase space distribution of electrons in the low-energy realm (up to several

MeV) of conventional accelerators. The technique can potentially serve for the attosecond compression of ultrabright multi-electron femtosecond pulses [32] leading to an enhancement of the peak electron density by a factor of $10^3$-$10^4$ and thus being of high interest for free-electron lasers or experiments requiring high quantum degeneracy of electron wavepackets.

**Acknowledgments**

The authors acknowledge funding from ERC grant "Near Field Atto", the Gordon and Betty Moore Foundation through Grant GBMF4744 "Accelerator on a Chip International Program – ACHIP" and BMBF via a project with contract number 05K16WEC.

**Figures:**

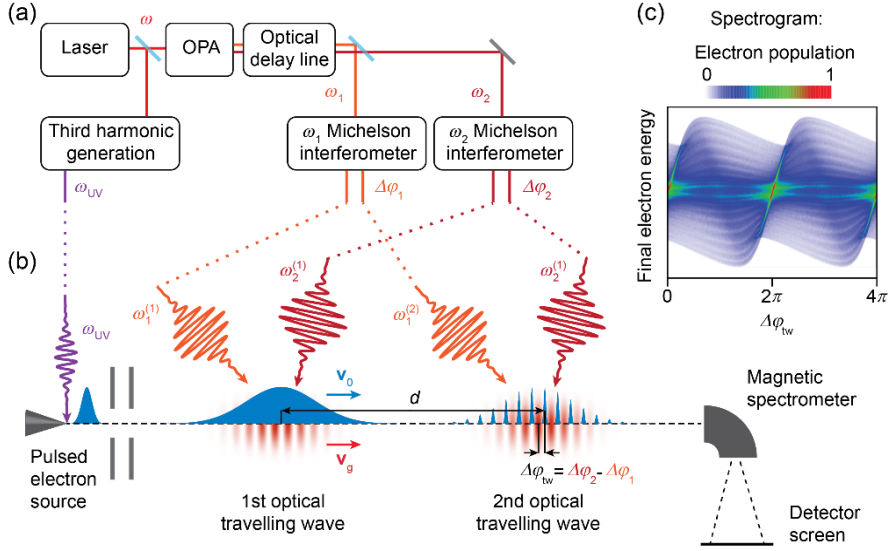

FIG. 1. Layout of the experimental setup for the generation and detection of attosecond electron pulse trains. (a) Laser setup serving for the simultaneous generation of the ultraviolet laser pulses for femtosecond pulsed electron emission ($\omega_{UV}$) and four infrared pulses ($\omega_1^{(1)}, \omega_2^{(1)}, \omega_1^{(2)}, \omega_2^{(2)}$) with controlled relative time delays for the generation of the two phase-locked optical travelling waves (further details of the optical setup are shown Fig. S1 in [30]). (b) Concept of the generation of the attosecond electron pulse train via the interaction of electrons with the first optical travelling wave and its characterization via the detection of electron spectra as a function of the relative phase $\Delta\varphi_{tw}$ between the first and the second travelling wave, separated by a drift distance $d$. (c) Simulated spectrogram showing the expected dependence of the final electron energy spectra on the relative phase difference $\Delta\varphi_{tw}$ between the two travelling waves.

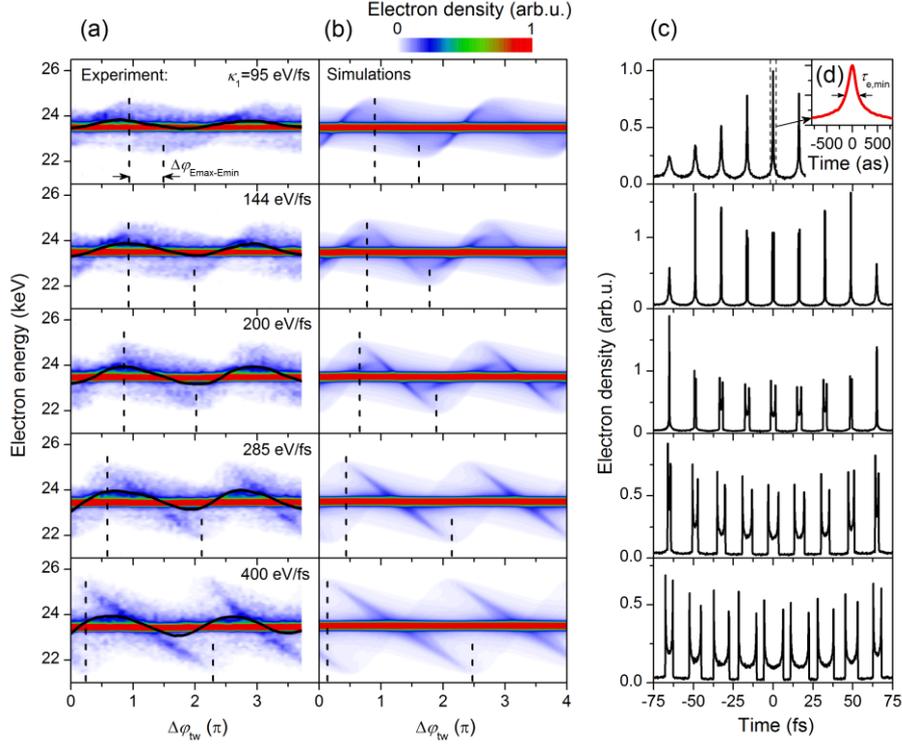

FIG. 2. Attosecond electron pulse train generation and characterization. (a) Measured spectrograms (electron spectra as a function of the relative phase $\Delta\varphi_{tw}$ between the two optical travelling waves, color coded) for different values of the linear chirp $\kappa^{(1)}$ introduced in the first interaction region. The drift distance between the two interaction regions is $d=41\pm2$ μm (Another set of data for $d=73\pm3$ μm is shown in Fig. S3 in [30]). The black curves show the center of mass of the electron energy spectra for interacting electrons. The electrons within the central peak at the initial energy are omitted for center of mass calculation. (b) Spectrograms (color coded) simulated using the measured values of $\kappa^{(1)}$ and $\kappa^{(2)}$ and the drift distance $d$. The dashed lines in (a) and (b) indicate the phase of energy maxima and minima for the evaluation of the phase shift $\Delta\varphi_{E\text{max-}E\text{min}}$. (c) Temporal evolution of the electron density in the center of the second interaction region corresponding to the calculated spectrograms shown in (b). For $\kappa^{(1)}=95$ eV/fs, the resulting pulse train consists of attosecond pulses with the shortest duration of $\tau_{e,\text{min}}=260$ as (FWHM) for the central pulse shown in detail in (d).

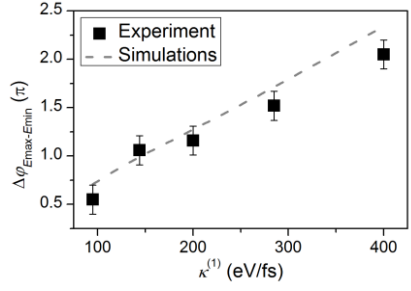

FIG. 3. Rotation of the electron distribution in the longitudinal phase space. Phase shift $\Delta\varphi_{E\text{max}-E\text{min}}$ between energy minima and maxima of the measured (squares) and the calculated (dashed line) spectrograms as a function of the linear chirp $\kappa^{(1)}$ introduced during the first interaction.